\documentstyle[aps,amsfonts,amssymb,twocolumn,epsfig]{revtex} 

\begin{document}
\draft

\wideabs{

\title{Electrical detection of spin accumulation and spin precession at room temperature in metallic spin valves}

\author{F.J. Jedema, M.V. Costache, H.B. Heersche,J.J.A. Baselmans, B.J. van Wees}
\address{Department of Applied Physics and Materials Science Center, University
of Groningen,\\ Nijenborgh 4, 9747 AG Groningen, The Netherlands
\\}
\date{\today}

\maketitle

\begin{abstract}

We have fabricated a multi terminal lateral mesoscopic metallic spin valve
demonstrating spin precession at room temperature, using tunnel barriers in
combination with metallic ferromagnetic electrodes as spin injector and
detector. The observed modulation of the output signal due to the spin
precession is discussed and explained in terms of a time of flight experiment
of electrons in a diffusive conductor. The obtained spin relaxation length
$\lambda_{sf}=500$ nm in an Al strip will make possible detailed studies of
spin dependent transport phenomena and allow to explore the possibilities of
the electron spin for new electronic applications at RT.
\end{abstract}}

A new direction is emerging in the field of
spintronics\cite{Spinbook,Wolfs2001,Jedema2001,Jedema2002}, where one wants to
inject spin currents, transfer and manipulate the spin information, and detect
the resulting spin polarization in nonmagnetic metals and semiconductors. A
first and successful attempt to electrically inject and detect spins in metals
dates back to 1985 when Johnson and Silsbee successfully demonstrated spin
accumulation in a single crystal aluminium bar up to temperatures of 77
K.\cite{Johnson1985,Johnson1988} In their pioneering experiments they were able
to observe spin precession of the induced non-equilibrium magnetization.
However, the measured signals were extremely small (in the pV range), due to
the relatively large sample dimensions as compared to contemporary technology.

In this letter we report spin precession in a diffusive Al strip at RT. The use
of tunnel barriers at the ferromagnetic metal-nonmagnetic metal (F/I/N)
interface and the reduced sample dimensions by 3 orders of magnitude, has
increased the output signal (V/I) of our device by 6 orders of magnitude as
compared to ref. \onlinecite{Johnson1985}. We find a spin relation length
$\lambda_{sf}=500$ nm in the Al strip at RT, which is within a factor of $2$ of
the maximal obtainable spin relaxation length at RT, being limited by the
inelastic phonon scattering processes. At lower temperatures larger spin
relaxation lengths can be obtained by reducing the impurity scattering rate, as
was previously reported.\cite{Jedema2002,Johnson1985,Johnson1988}


The samples are fabricated by means of a suspended shadow mask evaporation
process \cite{Dolan1977,trilayer} and using electron beam lithography for
patterning. The shadow mask is made from a tri-layer consisting of a $1.2~\mu
m$  PMMA-MA base layer (Allresist GMBH ARP 680.10 in methoxy-ethanol), a $40$
nm thick germanium (Ge) layer and on top a $200$ nm thick PMMA layer (Allresist
GMBH ARP 671.04 in chlorobenzene). The base and top resist layers have
different sensitivities for e-beam radiation, which enables a selective
exposure by varying the induced charge dose ($400~\mu C/cm^2$: both layers,
$100~\mu C/cm^2$: base layer) by the e-beam.

\begin{figure}[!thb]
\centerline{\psfig{figure=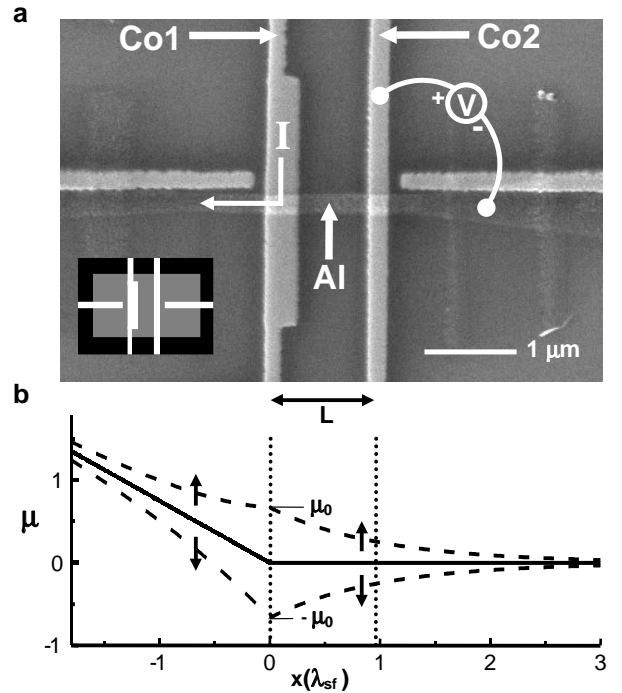,width=8 cm,clip=}} \caption{(a)
SEM-picture of the spin valve device. The current I is injected from Co1 into
the Al strip (left side) and the voltage V is measured between Co2 and the Al
strip (right side). Inset: center of the tri-layer shadow mask (see text).
Black: PMMA-MA/Ge bilayer. White: SiO$_2$ substrate. Grey: suspended Ge layer.
(b) The spatial dependence of the spin-up and spin-down electrochemical
potentials (dashed) in the Al strip. The solid lines indicate the
electrochemical potential (voltage) of the electrons in the absence of spin
injection.} \label{sample}
\end{figure}

In the first development step the top layer is developed, followed by a
anisotropic CF$_4$ dry etching to remove the exposed germanium layer. In a
third (wet) development step the PMMA-MA base layer underneath the Ge layer is
developed resulting in a suspended shadow mask, see the inset of Fig.
\ref{sample}a.

In a last step, the top resist layer is etched away by using an oxygen plasma.
After completion of the mask, a two step shadow evaporation procedure is used
to make the sample. First we deposit an Al layer from the left and right side
(see inset Fig. \ref{sample}a) under an angle of 25$^\circ$ with the substrate
surface at a pressure of 10$^{-6}$ mbar, thus forming a continuous Al strip
underneath the suspended Ge mask with a thickness of 50 nm.

Next, without breaking the vacuum, an Al$_2$O$_3$ oxide layer is formed at the
Al surface due to a 10 minutes O$_2$ exposure at 5 x 10$^{-3}$ mbar. In a third
step, after the vacuum is recovered, a 50 nm thick Co film is deposited from
below (see inset Fig. \ref{sample}a) under an angle of 85$^\circ$ with the
substrate surface. In Fig. \ref{sample}a a SEM picture is shown of a sample
with a Co electrode spacing of $L=1100$ nm. The conductivity of the Al and Co
strips were determined to be $\sigma_{Al}=~1.3\cdot 10^7 ~ \Omega ^{-1} m^{-1}$
and $\sigma_{Co}=~4.1\cdot 10^6 ~ \Omega ^{-1} m^{-1}$, whereas the resistance
of the Al/Al$_2$O$_3$/Co tunnel barriers were determined to be $800~\Omega$ for
the Co1 electrode and $2000~\Omega$ for the Co2 electrode at RT.


\begin{figure}
\centerline{\psfig{figure=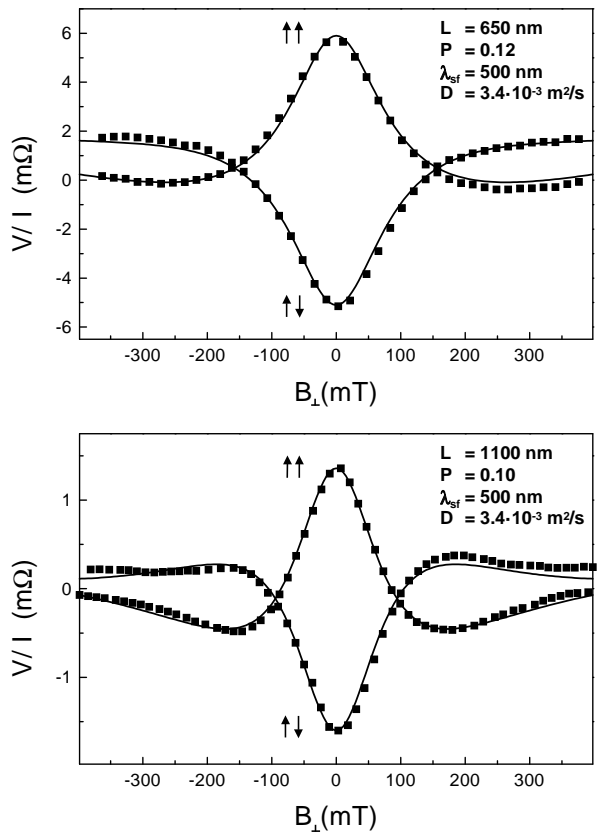,width=8 cm,clip=}} \caption{Modulation of
the output signal (V/I) due to spin precession as a function of a perpendicular
magnetic field $B\bot$, for L=650 nm and L=1100 nm. The solid squares represent
data taken at RT, whereas the solid lines represent the best fits based on eq.
\ref{precess}. We note that the fits incorporate the effect of a slight tilting
of the magnetization direction of the Co electrodes out of the substrate plane,
see Ref. \ref{jedema2002}.} \label{data}
\end{figure}

In our experiment we inject a spin polarized current ($I=100~\mu A$) from the
Co1 electrode via a tunnel barrier into the Al strip. The spin polarization $P$
of the current is determined by the ratio of the different spin-up and
spin-down tunnel barrier resistances $R^{TB}_\uparrow$ and $R^{TB}_\downarrow$,
which in first order can be written as
$P=(N_\uparrow-N_\downarrow)/(N_\uparrow+N_\downarrow)$.\cite{Julliere1975}
Here $N_\uparrow (N_\downarrow)$ is the spin-up (spin-down) density of states
at the Fermi level of the electrons in the Co electrodes. The unequal spin-up
and down currents cause the electrochemical potentials (densities)
$\mu_\uparrow,\mu_\downarrow$ of the spin-up and spin down electrons in the Al
strip to become unequal, see Fig. \ref{sample}b. The spatial dependence of
$\mu_\uparrow,\mu_\downarrow$ can be calculated by solving the 1-dimensional
spin coupled diffusion equations in the Al strip.\cite{son1,valet1} For
$x\geqq0$, we obtain:

\begin{equation}
\mu(x)_{\uparrow}=\mu_0\exp(\frac{-x}{\lambda_{sf}})~~\textrm{and}~\mu(x)_{\downarrow}=
-\mu_0\exp(\frac{-x}{\lambda_{sf}}) \;, \label{muup}
\end{equation}

where $\mu_0=\frac{eI\lambda_{sf}P}{2A\sigma_N}$ and
$\lambda_{sf}=\sqrt{D\tau_{sf}}$, $D$, $\tau_{sf}$, $\sigma_N$ and $A$ are the
spin relaxation length, diffusion constant, spin relaxation time, conductivity
and cross sectional area of the Al strip.

At a distance $L$ from the Co1 electrode the induced spin accumulation
($\mu_\uparrow-\mu_\downarrow$) in the Al strip can be detected by a second Co2
electrode via a tunnel barrier. The detected potential is a weighted average of
$\mu_\uparrow$ and $\mu_\downarrow$ due to the spin dependent tunnel barrier
resistances:

\begin{equation}
\mu_d=\frac{\pm P(\mu_\uparrow-\mu_\downarrow)}
{2}+\frac{(\mu_\uparrow+\mu_\downarrow)}{2} \;,\label{PotDet}
\end{equation}

where the + (-) sign corresponds with a parallel (anti-parallel) magnetization
configuration the Co electrodes. Using eqs. \ref{muup} and \ref{PotDet} we can
calculate the magnitude of the output signal (V/I) of the Co2 electrode
relative to the Al voltage probe at distance $L$ from Co1:

\begin{equation}
\frac{V}{I}=\frac{\mu_d-\mu_N}{eI}=\pm \frac{P^2 \lambda_{sf}}{2 A \sigma_N}
\exp(\frac{-L}{\lambda_{sf}})\;, \label{eq_spin_valve}
\end{equation}

where $\mu_N=(\mu_\uparrow+\mu_\downarrow)/2$ is the measured potential of the
Al voltage probe. Equation \ref{eq_spin_valve} shows that in absence of a
magnetic field the output signal decays exponentially as a function of
$L$.\cite{Jedema2002}

However, in the experiment the injected electron spins in the Al strip are
exposed to a magnetic field $B\bot$, directed perpendicular to the substrate
plane and the initial direction of the injected spins being parallel to the
long axes of Co electrodes. Because $B\bot$ alters the spin direction of the
injected spins by an angle $\phi=\omega_Lt$ and the Co2 electrode detects their
projection onto its own magnetization direction (0 or $\pi$), the spin
accumulation signal will be modulated. Here $\omega_L=g\mu_BB\bot/\hbar$ is the
Larmor frequency, g is the g-factor of the electron (~2 for Al), $\mu_B$ is the
Bohr magneton, $\hbar$ is Planck's constant divided by $2\pi$ and $t$ is the
diffusion time between Co1 and Co2. The observed modulation of the output
signal as a function of $B\bot$ at RT is shown in Fig. \ref{data}.

For a parallel $\uparrow \uparrow$ (anti-parallel $\uparrow \downarrow$)
configuration we observe an initial positive (negative) signal, which drops in
amplitude as $B\bot$ is increased from zero field. This is called the Hanle
effect in Refs. \cite{Johnson1985,Johnson1988}. The parallel and anti-parallel
curves cross each other where the average angle of precession is about 90
degrees and the output signal is close to zero. As $B\bot$ is increased beyond
this field, we observe that the output signal changes sign and reaches a
minimum (maximum) when the average angle of precession is about 180 degrees,
thereby effectively converting the injected spin-up population into a spin-down
and vice versa. We have fitted the data with eq. \ref{precess} and using
$\lambda_{sf}=500$ nm, $\sigma_N=e^2ND$  and $N=2.4\cdot10^{28}$
states/eV/$m^3$\cite{papa1986} we find the spin relaxation time $\tau_{sf}=65$
ps in the Al strip at RT to be in good agreement with theory.\cite{Fabian1999}
We note that half of the momentum scattering processes at RT is due to phonon
scattering, which implies that the spin relaxation length can be maximally
improved by a factor of $2$. A detailed discussion about spin relaxation times
is given in \onlinecite{jedema}.


\begin{figure}
\centerline{\psfig{figure=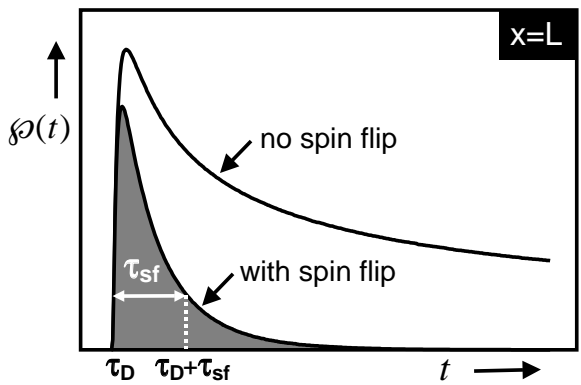,width=8 cm,clip=}} \caption{Probability
per unit volume that, once an electron is injected, will be present at $x=L$
without spin flip ($\wp(t)$) and with spin flip ($\wp(t)\cdot
exp(-t/\tau_{sf}))$, as a function of the diffusion time $t$. }
\label{distribution}
\end{figure}

Figure \ref{data} shows the amplitude of the oscillating output signal decays
with increasing $B\bot$, caused by the diffusive nature of the Al strip. In an
(infinite) diffusive 1D conductor the diffusion time $t$ from Co1 to Co2 has a
broad distribution $\wp(t)=\sqrt{1/4\pi Dt}\cdot exp(-L^2/4Dt)$, where $\wp(t)$
is proportional to the number of electrons per unit volume that, once injected
at the Co1 electrode (x=0), will be present at at the Co2 electrode ($x=L$)
after a diffusion time $t$. Therefore the output signal ($V/I$) is a summation
of all contributions of the electron spins over all diffusion times $t$:

\begin{equation}
\frac{V(B\bot)}{I}=\pm \frac{P^2}{e^2 N(E_F)A}\int_0^\infty \wp
(t)\cos(\omega_l t)\exp\left(\frac{-t}{\tau_{sf}}\right)dt\;. \label{precess}
\end{equation}

In Fig. \ref{distribution} $\wp(t)$ is plotted as a function of $t$, showing
that the integral of $\int_0^\infty\wp(t)dt$ is diverging. So even when
$\tau_{sf}$ is infinite the broadening of diffusion times will destroy the spin
coherence of the electrons present at Co2 and hence will lead to a decay of the
output signal. However, a sign reversal of the output signal is still observed
because only the electrons present at Co2 carrying their spin information are
relevant. The exponential factor in the integral of eq. \ref{precess},
describing the effect of the spin flip scattering, will cut off the diffusive
broadening of $\wp(t)$ and create a window of diffusion times from $\tau_D$ to
$\tau_D+\tau_{sf}$, see Fig. \ref{distribution}. Here $\tau_D=L^2/2D$ is the
diffusion time corresponding to the peak of $\wp(t)$ in absence of spin flip
scattering. The condition to observe more than a half period of modulation
imposes $\phi_{ave}=\omega_L\tau_D \geq \pi$, whereas a limitation on the
diffuse broadening imposes the condition $\Delta\phi=\omega_L\tau_{sf}\leq\pi$.
Using $\tau_D=L^2/2D$ we find with this simplified picture that the requirement
in order to observe at least half a period of oscillation is approximately
given by: $L\geq \sqrt{2}~\lambda_{sf}$.

Using the program Mathematica we can solve the integral $Int(B\bot) =
\int_0^\infty \wp (t)\cos(\omega_l t)\exp\left(\frac{-t}{\tau_{sf}}\right)dt$
and we find:

\begin{equation}
Int(B\bot) =
Re(\frac{1}{2\sqrt{D}}\frac{\exp\left[-L\sqrt{\frac{1}{D\tau_{sf}}-i
\frac{\omega_L}{D}}\right]}{\sqrt{\frac{1}{\tau_{sf}}-i\omega_L}}) \;.
\label{intB}
\end{equation}

Equation \ref{intB} shows that in the absence of precession ($B\bot=0$) the
exponential decay of eq. \ref{eq_spin_valve} is recovered. It can be shown by
using standard goniometric relations that eq. \ref{intB} is identical to the
solution describing spin precession obtained by solving the Bloch equations
with a diffusion term.\cite{Johnson1988} In particular we find
$Int(B\bot)=\frac{1}{2}\frac{\sqrt{\tau_{sf}}}{2D}F_1\{b,l\}$, where
$F_1\{b,l\}$ is derived in Ref. \onlinecite{Johnson1988},
$b\equiv\omega_L\tau_{sf}$ is the reduced magnetic field parameter and
$l\equiv\sqrt{\frac{L^2}{2D\tau_{sf}}}$ is the reduced injector-detector
separation parameter.

To conclude, we have demonstrated spin precession in an Al strip at RT. As a
final note we believe that our obtained value $P\thickapprox10\%$\cite{Tedrow1}
is too low and we anticipate that the output signal of our device can be
improved by more than an order of magnitude by improving the material
properties of the Co material.\cite{Egelhoff1} The authors wish to thank the
Stichting Fundamenteel Onderzoek der Materie (FOM) and NEDO (project
'Nano-scale control of magnetoelectronics for device applications') for
financial support.

\end{document}